\documentclass{Rinton-P9x6}

\usepackage{amsmath}
\usepackage{amssymb}
\usepackage{overcite}

\newcommand{\pdo}[1]{\ensuremath{\frac{\partial }
        {\partial #1 }}}

\newcommand{\slashletter}[1]{\ensuremath{\kern+0.1em /\kern-0.65em #1}}

\begin{document}

\title{Fermion Determinants 2003}

\author{M.P. Fry}
\address{School of Mathematics, University of Dublin, Dublin 2, Ireland}

\maketitle

\abstracts{It is recommended that lattice QCD representations of the
fermion determinant, including the discretization of the Dirac
operator, be checked in the continuum limit against known QED
determinant results.  Recent work on the massive QED fermion
determinant in two dimensions is reviewed. A feasible approach to the
four-dimensional QED determinant with $O(2) \times O(3)$ symmetric
background fields is briefly discussed.}

\section{Introduction}

The calculation of \emph{every} physical process requires the gauge
field measure

\begin{equation}
\label{Eq_01}
d\mu(A) = \frac{1}{Z}
  e^{-[(1/4) \int d^4x \text{Tr} F_{\mu\nu} F_{\mu\nu}]}
  \text{det}(\slashletter{D} + m) \, \Pi \, dA^{a}_{\mu}(x),
\end{equation}

\noindent
obtained by making the Wick rotation to Euclidean space and integrating
out the fermion fields. Here
$D_{\mu} = \frac{1}{i} \partial_{\mu} - g T^{a} A^{a}_{\mu}$,
where $T$ is some real representation of a compact gauge group, and
$Z$ is the partition function. Granted its fundamental importance, it
is curious that the fermion determinant, $\text{det}(\slashletter{D} + m)$,
has generally failed to engage the attention of physicists
outside lattice QCD since the 1970s and early 80s. Indeed, until
recently~\cite{A_Fry03} it was not even known what the strong coupling
limit of $det$ with $m \neq 0$ in $\text{QED}_2$ was for a general class
of centrally symmetric background fields $F_{\mu\nu}$, though the
question was asked twenty five years ago.~\cite{A_Balian78} Part of
the problem is that $det$ is a nonlocal function of
$F_{\mu\nu}$. Moreover, it has to be known for general $F_{\mu\nu}$ to
say anything useful about the above measure.  Consequently, $det$ is
hard to calculate and so before the advent of large machines $det$ was
expanded in a power series in $g$. Now there is nothing wrong with
this provided the remainder after N terms,
$|\text{det} - \sum^{N}_{n=0} \text{det}_{n} g^{n}|$, can be
bounded. This requires nonperturbative information on $det$ that
lattice versions can provide with $\gtrsim 500$ GFLOP machines. Several
lattice groups are now engaged in this endeavor.~\cite{Collabs}

The following has to be kept firmly in mind: lattice representations of
$det$, which are taken to include the many lattice discretizations of
the Dirac operator now in use, should be tested against known
infinite-volume, continuum results for the fermion determinant to
estimate any associated systematic computational error. Aside from
such general results as the positivity of $det$ and the diamagnetic
bound $det \leq 1$ in two and three
dimensions~\cite{A_Brydges79,B_Seiler81} there is little to test $det$
against in the physical case $m \neq 0$. Therefore, we advocate the
following remedy: test computations of $det$ against known continuum
results for Abelian background fields in two, three, and four
dimensions. This would quantify discretization errors precisely for a
particular representation of $det$. Work in this direction has already
begun.~\cite{A_Chiu98}

A survey of results for fermion determinants was presented at this
workshop in 2001~\cite{A_Fry02} and will not be repeated
here. Instead, we will comment on some recent results found
in~\cite{A_Fry03}. We will conclude with a brief discussion of how
progress can be made on the strong coupling limit of the fermion
determinant in $\text{QED}_4$ for four-variable background fields.

\section{Massive $\text{QED}_2$}

The one-loop effective Euclidean action $S_{\text{eff}} = -lndet$ is
calculated from the ratio
$\text{det}(\slashletter{P} - e\slashletter{A} + m)/det(\slashletter{P} + m)$
of Fredholm determinants of Dirac operators. This ratio is formal and
mathematical sense has to be made of it, such as Schwinger's proper
time definition.~\cite{A_Schwinger51} Starting from this definition a
new representation of $lndet$ for massive $\text{QED}_2$ was
obtained~\cite{A_Fry93}, namely

\begin{multline}
\label{Eq_02}
\pdo{e} lndet = \frac{e}{\pi} \int d^2r \, \varphi \; \partial^2\varphi\\
  + 2m^2 \int d^2r \, \varphi(\mathbf{r})
  < \mathbf{r} | ( H_{+} + m^2)^{-1} - ( H_{-} + m^2)^{-1} | \mathbf{r} >,
\end{multline}

\noindent
where the supersymmetric operator pair
$H_{\pm} = (\mathbf{P} - e\mathbf{A})^2 \mp eB$ are obtained from the
Pauli operator
$(\mathbf{P} - e\mathbf{A})^2 - \sigma_3 eB$,
$A_{\mu} = \epsilon_{\mu\nu} \partial_{\nu} \phi$ and
$B = - \partial^2 \phi$.
The first term on the right-hand side of (\ref{Eq_02}) is the
contribution to the massive determinant from the massless Schwinger
model.~\cite{A_Schwinger62} By specializing to centrally symmetric,
square-integrable magnetic fields with range $R$, (\ref{Eq_02}) can be
put into the form of a partial wave expansion:~\cite{A_Fry03}

\begin{multline}
\label{Eq_03}
\pdo{e} ln det = \frac{e}{\pi} \int d^2r \phi \partial^2 \phi\\
  - 2m^2 \sum^{\infty}_{l=-\infty} \int^{R}_{0} dr r
    \left(
      G^{+}_{l}(r, r; me^{\frac{i\pi}{2}})
      - G^{-}_{l}(r, r; me^{\frac{i\pi}{2}})
    \right)
    \phi(r)\\
  + \frac{im^2}{\pi} \sum^{\infty}_{l=-\infty} e^{-i\pi|l|}
    \left(
    e^{2i\delta^{+}_{l}\left(e^{\frac{i\pi}{2}}m\right)}
    - e^{2i\delta^{-}_{l}\left(e^{\frac{i\pi}{2}}m\right)}
    \right)\\
  \times
  \int^{\infty}_{R} dr \, r \ln \left(\frac{r}{R}\right)
    K^2_{|\frac{e\phi}{2\pi} - l|} (mr).
\end{multline}

\noindent
Here $G^{\pm}_l$ are outgoing-wave Green's functions for $H_{\pm,l}$,
$\delta^{\pm}_{l}$ are the positive/negative chirality partial wave
phase shifts, and $\Phi$ is the total flux of $B$. Both $G^{\pm}_{l}$
and $\delta^{\pm}_{l}$ are analytically continued into the upper half
k-plane by setting $k = me^{i\pi/2}$. The representation (\ref{Eq_03})
is exact. In order to penetrate deeply into the nonperturbative regime
we considered the limit $mR << 1$ followed by $|e\Phi| >> 1$. This is
possible because the zero modes of $H_{\pm,l}$ and their threshold
resonance states can be calculated explicitly. For $B(r) > 0$ with the
two alternative sets of boundary conditions $B(R) = 0$ or
$\lim_{r \to R^{-}} B'(r) < 0$ with $B(R) > 0$, the result in both
cases is

\begin{equation}
\label{Eq_04}
\lim_{|e\Phi| >> 1} \lim_{mR << 1} lndet
  = -\frac{|e\Phi|}{4\pi} \ln\left(\frac{|e\Phi|}{(mR)^2}\right)
    + O(|e\Phi|,(mR)^2|e\Phi|\ln(|e\Phi|)).
\end{equation}

\noindent
The analysis in \cite{A_Fry03} of the remainder terms in (\ref{Eq_04})
is not yet sharp enough to exclude the logarithm factor.

Several comments are in order. Firstly, we see that the presence of
mass profoundly modifies the determinant. If $m = 0$ \emph{ab initio},
then the second term in (\ref{Eq_02}) is absent and Schwinger's result
is regained.  But for $m \neq 0$ and $|e\Phi| >> 1$ there is a build
up of zero modes, and these eventually cancel the Schwinger
contribution to $lndet$. In fact, the factor $|e\Phi|/4\pi$ in
(\ref{Eq_04}) is proportional to the number of zero modes.

Secondly, the minus sign in (\ref{Eq_04}) is a reflection of the
paramagnetism of charged fermions in a magnetic field and is
consistent with the diamagnetic bound 
$det \leq 1$.~\cite{A_Brydges79,B_Seiler81,A_Weingarten80}

Thirdly, there is a logarithmic dependence on $e\Phi$. This is not
understood. The $\text{QED}_2$ determinant is an entire function of
$e$ of order 2 for the class of fields
considered.~\cite{B_Seiler81,A_Seiler80} The zeros of
$det_{\text{QED2}}$ are spread over the complex $e$-plane in quartets
or in imaginary pairs. For $e^2 \to \infty$ the zeros of
$det_\text{QED2}$ conspire to produce the logarithmic dependence on
$e\Phi$ seen in (\ref{Eq_04}) so that the maximal growth of
$det_{\text{QED2}}$ is somewhere off the real axis of the complex
$e$-plane.  What is the physics behind this?

Finally, although the calculation leading to (\ref{Eq_04}) was for a
special class of magnetic fields the result only depends on a global
property of $B$, namely $\Phi$. So it is possible that it also holds
for noncentral, square-integrable magnetic fields. We know that the
zero modes dominate in the limit considered and that their number only
depends on $e\Phi$ for any reasonable magnetic field.~\cite{A_Aharonov79}

\section{Duality}

The continuation of a fermion determinant back to Minkowski space
involves continuing the background field as well. This is
nontrivial. In Euclidean space the calculation of $det_{\text{QED2}}$
for centrally symmetric magnetic fields with rapid falloff reduces to
a scattering problem with outgoing wave boundary conditions. Going
back to Minkowski space and transforming the magnetic field to a
one-dimensional electric pulse turns the problem into one of pair
production with entirely different boundary conditions. This has been
carefully investigated for the exactly solvable single-variable
magnetic field $B(x) = B sech^2 ( x / \lambda )$~\cite{A_Dunne98} and for
more general one-variable fields using a WKB
approach.~\cite{A_Dunne99} Assuming that the Wick rotation goes
through in the two-variable case for the fields considered in the
derivation of (\ref{Eq_04}) we obtained~\cite{A_Fry03}

\begin{equation}
\label{Eq_05}
2\pi \pdo{m^2} S^{\text{QED}_4}_{\text{eff}}(E)
  = i L_1 L_2 \, lndet_{\text{QED}_2} (B \to e^{-i\pi/2} E)\\
    + \frac{L_1 L_2 ||E||^2 e^2}{12 \pi m^2}.
\end{equation}

\noindent
Here $S^{\text{QED4}}_{\text{eff}}$ is the $\text{QED}_4$ one-loop
Minkowski metric effective action obtained by making the duality
transformation that takes the static magnetic field
$B (x_1, x_2) \, \hat{\mathbf{k}}$ to the functionally equivalent
electric pulse $\mathbf{E} \, (x_3, t) = B (x_3, t) \,
\hat{\mathbf{k}}$. $L_1$ and $L_2$ are the sides of the space box in
the $x_1$ and $x_2$ directions. The rules for making the analytic
continuation we have outlined are found in Sec. VI of
\cite{A_Fry03}.

Inserting the result (\ref{Eq_04}) into the right-hand side of
(\ref{Eq_05}) gives the result

\begin{equation}
\pdo{m^2} \text{Im} S^{\text{QED}_4}_{\text{eff}} < 0,
\end{equation}

\noindent
indicating that the pair production probability in the presence of $E$
decreases with increasing fermion mass. This physically reasonable
result depends on the minus sign in (\ref{Eq_04}), which we have noted is a
reflection of the Euclidean diamagnetic bound $det \leq 1$ in two
dimensions. This bound is difficult to establish. It is first obtained
on a lattice. Then the continuum limit has to be taken, and finally
it has to be shown that the determinant so defined agrees with what
physicists regard a determinant to be, such as Schwinger's proper
time definition. It is reassuring that the diamagnetic bound has an
immediate physical interpretation in Minkowski space.

Realistic laboratory electric pulses require the calculation of $det$
for more general magnetic fields than those considered so far. Then
the question arises as to whether the Wick rotation to the associated
dual electric pulse goes through. The answer should give us a better
physical understanding of $det$.

\section{$\text{QED}_4$}

There are some major results. For example, if
$A_{\mu} \in L^{p}(\mathbb{R}^4)$, $p > 4$, then $det_{\text{QED4}}$
can be expressed as a renormalized determinant that is an entire
functin of $e$ of order 4 for massive fermions, even though
$F_{\mu\nu}$ may not be square integrable. This is summarized in
reference \cite{A_Seiler80} which refers the reader to the original
sources. By an easy generalization of the two-dimensional case
discussed in \cite{A_Seiler80} to four dimensions the same result
holds for massless fermions provided
$A_{\mu} \in L^{p}(\mathbb{R}^4)$ for some $p$ in the open interval
$(\frac{4}{3},4)$. If $A_{\mu}$ is ``winding'', i.e., falls off as
$1/r$, and the fermions are massless then the renormalized determinant
defined in reference \cite{A_Seiler80} ceases to exist, necessitating
some other definition of the determinant. For massive fermions and
$A_{\mu}$ winding the renormalized determinant exists but requires
knowledge of the Dirac operator's zero modes. These results also hold
for $\text{QCD}_4$. For the special case of background fields with
$O(2) \times O(3)$ symmetry the author has recently been able to
calculate all the zero modes explicitly in $\text{QED}_4$. The
symmetry $O(2) \times O(3)$ reduces the problem of calculating
$det_{\text{QED4}}$ to a centrally symmetric scattering problem. The
author sees no barrier to results for such fields in the near future.

\end{document}